\documentclass[10pt]{article}
\usepackage{amsthm,amssymb,amsmath,epsfig, graphicx, fullpage,mathrsfs}

\newcommand{\p}{\partial}
\renewcommand{\vec}[1]{\mathbf{#1}}
\newcommand{\abs}[1]{\left|#1\right|}
\newcommand{\ept}[1]{\left\langle#1\right\rangle}
\newcommand{\rin}{r_{\textrm{in}}}
\newcommand{\rout}{r_{\textrm{out}}}
\newcommand{\zlow}{z_{\textrm{low}}}
\newcommand{\zhigh}{z_{\textrm{high}}}

\newcommand{\vw}{v_w}
\newcommand{\xb}{\vec{x}}
\newcommand{\rb}{\vec{r}}
\newcommand{\ub}{\vec{u}}
\newcommand{\Db}{\vec{D}}
\newcommand{\Fb}{\vec{F}}
\newcommand{\Eb}{\vec{E}}
\newcommand{\Tb}{\vec{T}}
\newcommand{\ds}{\hat{\vec{d}}_s}
\newcommand{\del}{\mbox{\boldmath{$\nabla$}}}

\DeclareMathOperator{\tr}{tr}

\title{The Stochastic Flow Rule:\\ A Multi-Scale Model for Granular Plasticity}
\author{Ken Kamrin, Chris H. Rycroft, and Martin Z. Bazant \\ Department of Mathematics  \\ Massachusetts Institute of Technology  \\ Cambridge, MA 01239 }
\date{\today}

\begin{document}

\maketitle

\begin{abstract}
  In spite of many attempts to model dense granular flow, there is
  still no general theory capable of describing different types of
  flows, such as gravity-driven drainage in silos and wall-driven
  shear flows in Couette cells. Here, we summarize our recent proposal
  of the Stochastic Flow Rule (SFR), which is able to describe these
  cases in good agreement with experiments, and we focus on testing
  the theory in more detail against brute-force simulations with the
  discrete-element method (DEM). The SFR is a general rate-independent
  constitutive law for plastic flow, based on diffusing ``spots'' of
  fluidization. In the case of quasi-2D granular materials, we assume
  limit-state stresses from Mohr-Coulomb plasticity and postulate that
  spots undergo biased random walks along slip-lines, driven by local
  stress imbalances.  We compare analytical predictions of the SFR
  against DEM simulations for silos and Couette cells, carrying out 
  several parametric studies in the latter case, and find good
  agreement.
\end{abstract}

\section{Introduction}
Even after centuries of study, dense granular flow remains an enigma
for both physicists and engineers~\cite{jaeger96,aranson06}.  For
dilute, collisional granular media, the kinetic theory of dissipative
gases has been quite successful at describing the mean flow
statistics~\cite{campbell90}. On the opposite side of the spectrum,
where the material is solid-like and possesses a yield function, the
stresses and (to a lesser extent) the dynamics can be adequately
described by methods of modern soil
mechanics~\cite{nedderman,schofield}. However, the intermediate regime
where the flow is slow, steady, and dense, has resisted many
theoretical attempts~\cite{alltheories,kamrin06}.

We have recently proposed the Stochastic Flow Rule (SFR) for dense
granular flows \cite{kamrin06}. It is a rate-independent constitutive
law for steady flow which we believe naturally extends the
applicability of traditional soils plasticity. Since it is rooted in a
general theory for granular stresses, the SFR can be applied in any
environment with two-dimensional (2D) symmetry, a level of generality
not common in other models. Though the mean stress field is assumed to
obey a continuum law, the SFR does not, however, treat the \emph{flow}
as continuous, but rather as the superposition of many discrete
plastic flow events. The carriers of plastic flow are ``spots'' of
local fluidization, which diffuse through the material in response to
stress imbalances. The Spot Model provides a precise and general
mechanism for random-packing dynamics~\cite{bazant06}, which can be
either used as a basis for multiscale particle
simulations~\cite{rycroft06a} or averaged to obtain simple continuum
flow equations, which can be solved exactly. Here we take the latter
approach and compare the analytical predictions of the SFR to
brute-force simulations of dense granular flows by the Discrete
Element Method (DEM).

We begin by summarizing the theory, starting with the notion of the
``spot'', as a discrete carrier of granular motion. We then review
limit-state Mohr-Coulomb plasticity for stresses, and proceed to
describe how the local stress state biases the motion of spots. With
the SFR fully defined, we then solve for the mean flow field in the
draining-silo and Couette annular-flow geometries and compare to DEM
simulations fo both geometries. For annular flow in particular, we test
predictions of the dependence of the flow profile on the particle
contact friction, rotation rate of the inner wall, and inner wall
radius, which constitute rigorous tests of the theory.

\section{Theory}
\subsection{Spots and spot dynamics}
At the outset, we must first discuss a particular property of granular flows
underscoring their discrete nature. Unlike a true continuum, a steady flow
field for a granular media is actually only steady in a time-averaged sense.
There are constant microscopic fluctuations occurring in the flow caused
chiefly by geometric packing constraints: grains cannot pass through other
grains, so frequently the particle motion must fluctuate from the mean to
achieve flow.

In experiments and simulations of silo drainage, a pronounced
correlation length for these fluctuations was observed \cite{choi05,
  choi04, rycroft06a}. This length, typically $3d$ to $5d$, where $d$
is the particle diameter, implies that a grain does not move
independent of its surroundings. Rather, it travels in a collective
fashion, as if dragging a set of nearest neighbors with it as it
moves.

\begin{figure}[t]
\begin{center}
(a)
\epsfig{clip,width=1.6in, file=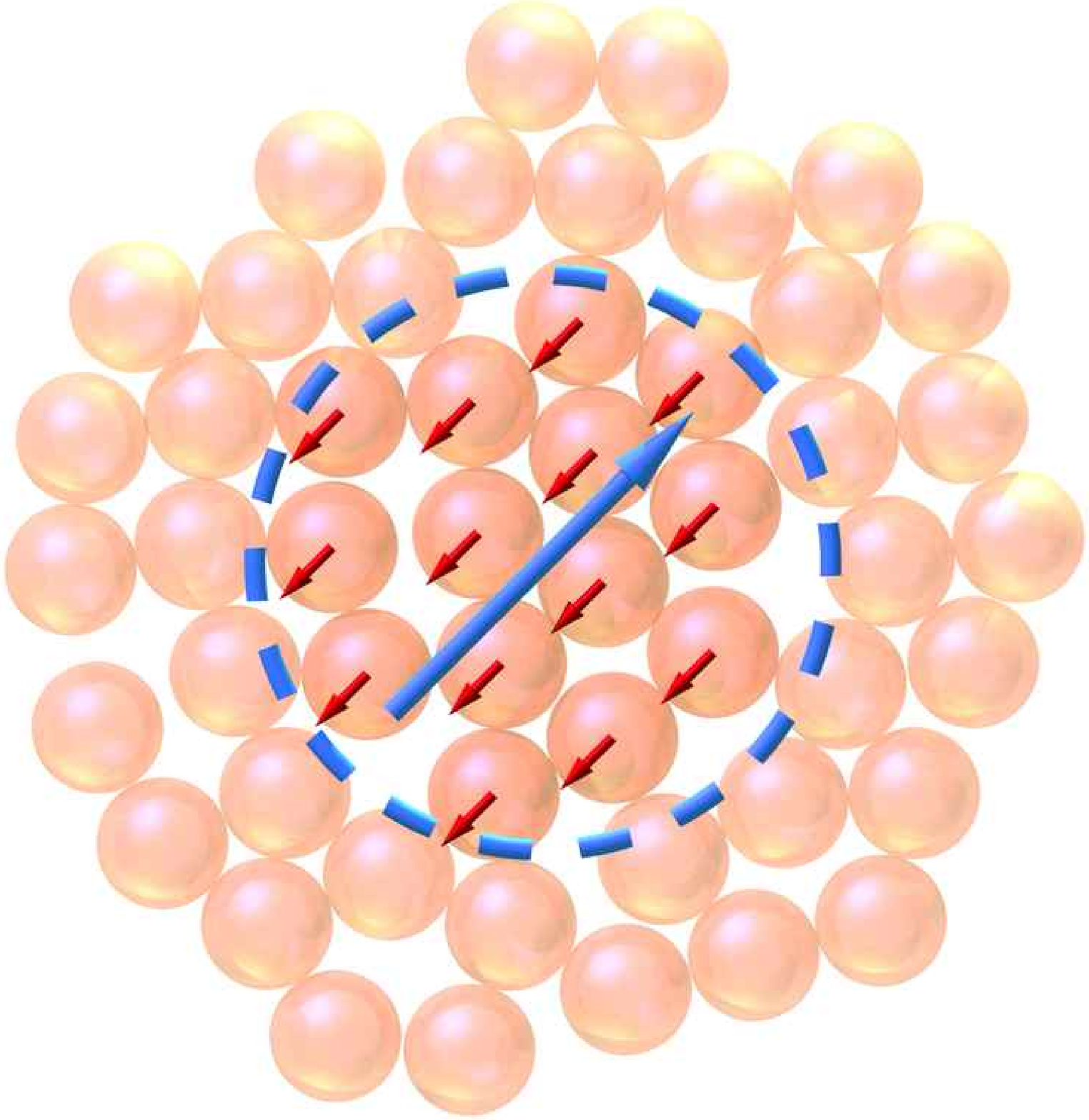} \qquad \qquad (b)
\epsfig{clip,width=2in, file=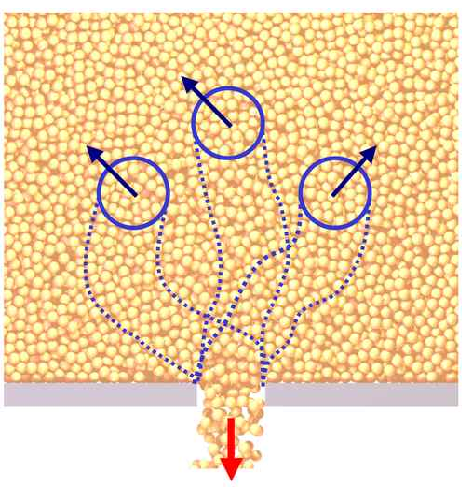}
\caption{Spots as carriers of plastic deformation in granular materials.
Cartoon of basic spot motion. A spot of local fluidization, carrying some free
volume, moves to the upper right causing a cooperative displacement of
particles, to the lower left, opposing the spot displacement. (b) In silo
drainage, spots are injected at the orifice and perform random walks biased
upward by gravity, causing downward motion of particles.}
\label{fig:spot}
\end{center}
\end{figure}

Based on this evidence for correlated motion, Bazant constructed the
Spot Model for random packing dynamics and applied it to the case of
granular drainage \cite{bazant06}. The model is based on a collective
microscopic mechanism for flow, where particles move to oppose the
random motion of ``spots'', as depicted in Figure \ref{fig:spot}. When
a spot displaces, the nearby affected particles respond by displacing
in a block-like fashion in the opposite direction. The spot can thus
be viewed as effectively carrying free volume and would typically be
associated with a slight increase in the local interstitial volume of the
packing. As a first approximation, identical spots undergo independent
random walks through the material, although interactions between spots
(e.g. carried by the stress field) and statistical variations in spot
size and shape could be considered as well.

Each step of its random trajectory, a spot chooses the displacement $\Delta
\xb_s$ following a biased random process. The mean displacement, which
is effectively the deterministic part of the motion, is given by
\begin{equation}\label{eq:driftdef}
  \ept{\Delta \xb_s}=\Delta t \Db_1
\end{equation}
for step duration $\Delta t$ and drift vector $\Db_1$. The random part
of the motion can be described at the simplest level by 
\begin{equation}\label{eq:diffdef}
  \ept{\Delta \xb_s \otimes \Delta \xb_s}= 2 \Delta t \Db_2
\end{equation}
where $\Db_2$ is the symmetric tensor of diffusion coefficients
(co-variance matrix) measuring the random fluctuations in each
direction. These coefficients suffice to construct a continuous
Fokker-Planck (drift-diffusion) equation to approximate the spot
concentration,
\begin{equation}
\frac{\partial \rho_s}{\partial t} + \del\cdot(\Db_1\rho_s) = 
(\del\otimes\del):(\Db_2\rho_s) + \ldots
\end{equation}
where omitted correction terms (in the Kramers-Moyall expansion)
depend on higher moments of the spot displacement~\cite{risken}.  For
example, in a simple model of silo drainage (ressembling the classical
kinematic~\cite{nedderman79} and void~\cite{lit58,mullins72} models),
the drift vector may be chosen to point uniformly upward, since
particles drift downward due to gravity, with a constant diffusivity
tensor to obtain spot trajectories as sketched in Figure
\ref{fig:spot}(b).

In steady-state flow, the spot concentration satisfies the
time-independent drift-diffusion equation,
\begin{equation}\label{eq:fokker}
\del\cdot (\Db_1\rho_s)=(\del \otimes \del):(\Db_2 \rho_s)
\end{equation}
where $\mathbf{A}:\mathbf{B}=A_{ij}B_{ij}$. Once solved, the mean particle
velocity field $\ub$ can be found by superposing the effects of all spots on
all the particles. This is easiest to see as a convolution of some influence
function $w(\rb,\rb')$, representing the amount of influence a spot centered at
$\rb'$ has on a particle at $\rb$, and the negative spot flux vector
\begin{equation}\label{eq:unconv}
\ub^*=- \Db_1\rho_s+\del\cdot (\Db_2 \rho_s).
\end{equation}
Thus the particle velocity $\ub$ is
\begin{equation}\label{eq:ub}
\ub=\int w(\rb,\rb')\ub^*(\rb')d\rb'.
\end{equation}
The influence function, which in simple language describes the spot's
shape, should have a characteristic width of three to five particle
diameters, so that spots match known correlation length data. In many
situations, $\ub^*$ is close to $\ub$ since the convolution with $w$
tends only to smooth out sharp changes in the spot flux density.

There are several ways to use the multiscale description of the Spot
Model, as have been demonstrated in the case of silo drainage.  Given
the drift field $\Db_1$, diffusion tensor field $\Db_2$, and for extra
clarity the spot influence function $w$, equations (\ref{eq:fokker})
and (\ref{eq:ub}) could be solved analytically or numerically to give
the mean flow field; the statistical trajectories of particles could
also be constructed, thus linking particle diffusion to the diffusion
of free volume~\cite{bazant06}. The microscopic basis for the Spot
Model, however, also allows for explicit multiscale simulations of all
particles in a granular flow, alternating between mesoscopic spot
displacements and cooperative particle motion; this approach yields
remarkably realistic flowing packings (compared to DEM simulations) in
silo drainage~\cite{rycroft06a}. This suggests that the spot mechanism
provides a robust description of how granular flow occurs and could
apply in a more universal sense, beyond just the gravity driven
drainage from a silo.

The difficult question, if we want to extend the spot model to an
arbitrary environment, is how can we derive the spot dynamics (or at least
$\Db_1$ and $\Db_2$) from mechanical principles? The intuition behind
the simple spot dynamics in silo drainage has no clear extension to
other situations, such as an annular Couette cell, where the flow is
unaffected by gravity.  Instead, the spot dynamics must be related to
stresses in the material, not only due to body forces, but also due to
forces transmitted from moving boundary tractions.

\subsection{Quasi-static granular stresses}

In a dense granular material, forces are transmitted across discrete
contact networks, which can be rather complicated and varying at the
same length scale as the velocity profile or even the container
geometry. Even the simplest question of how forces are transmitted in
static granular heaps has been the subject of considerable
controversy~\cite{bouchaud95,coppersmith96,mueth98,silbert02c,landry03}.
Nevertheless, it seems reasonable that a continuum theory could at
least describe the mean stress state in a steady flow as a starting
point for applying the SFR. Some of the random fluctuations in the
force network could also be manifested in the random motion of spots
coupled to mean stresses.

Most common continuum theories for solid-like material that can
plastically flow utilize some type of elasticity, generating the
stress tensor $\Tb$ from a small elastic strain $\Eb^e$ via
$\Tb=\mathscr{C}[\Eb^e]$ where the fourth rank tensor function
$\mathscr{C}$ is determined by the chosen elasticity
formulation. Elliptic systems of equations, such as the laws of linear
isotropic elasticity, have solutions for any given set of boundary
tractions or displacements, and thus complete boundary data must be
supplied.

This is problematic for granular materials since the boundary stresses
in a static assembly are frequently the quantity of interest. For
example, the pressure distribution under a sand-pile or the wall
stresses in a bin have both been studied extensively, as noted
above. Static granular boundary conditions typically develop
internally as a response to the construction proceedure and often
cannot be fluctuated without plastically deforming the assembly.

Hyperbolic laws of stresses have the benefit of constraint: if the tractions
along one boundary region are chosen, their effect can propagate through the
bulk to other boundaries and place constraints on the allowable tractions
there. Some non-linear elasticity laws have been shown to yield unique
solutions under restrained boundary information \cite{jiang03}, but it appears
a different continuum approach may be more naturally suited for granular
materials.

The traditional approach from soil mechanics is not a strain-based stress
formulation, but rather an analysis based more closely on a yield function. The
Coulomb yield function defines failure whenever $\abs{\tau/\sigma}=\mu=\tan\phi$
where $\tau$ is the shear stress on a plane, $\sigma$ the normal stress, and
$\mu$ is an internal friction coefficient unique to the material. This yield
function akin to a cohesionless dry friction law seems the natural choice for a
continuum description of granular materials---applying higher compression
locks the grains more securely, and thus makes it harder to shear them. Our
experience with granular piles also direct us to this yield function. If we
pour sand from a narrow spout onto a rough table, grains build up on the
surface of the sand cone until the cone angle reaches a certain value at which
point grains always slide down the surface, analogous to how a frictional block
resting on a board will start to slide if the board is tilted above some
critical angle.

In the Critical State Theory of soils \cite{schofield}, the stresses
throughout a moving soil eventually converge onto a ``critical state line''
where pressure $p$ and stress deviator $q$ are linearly related. This locus
strongly resembles the Drucker-Prager yield criterion, a simplified Coulomb
yield law for 3D. The deformation rate in such slow flows does not
significantly contribute to the stress state.

For this and other reasons we too will adopt a rate-independent formulation for
granular stresses, which should owe well to the slow flowing regime we are
primarily concerned with. To connect the flowing to the static regime, a common
assertion is that the yield criterion is being met at all points in the
material whether or not the material is flowing yet \cite{nedderman}. When
static material is brought to steady motion under this assertion, the locus of
static stresses already resembles the critical state line and thus no new
stress laws are needed.

Materials obeying this traditional hypothesis are said to be at
\emph{limit-state}. Let us focus entirely on 2D geometries (plane strain)
complete with a 2D stress tensor $\Tb$ defined in the plane of the flow. By
requiring the yield criterion to be met at all points, the limit-state
assumption implies the following constraint $\frac{1}{\sqrt{2}}\abs{\Tb_0}=\sin\phi \left(\frac{1}{2}\tr \Tb\right)$
where $\Tb_0$ is the deviatoric stress tensor and $\abs{\mathbf{A}}=
\sqrt{\mathbf{A}:\mathbf{A}}$. A simple way to uphold this constraint
is to rewrite the stress field in terms of two stress parameters (the
Sokolovskii variables~\cite{Sok}): $p$ the average pressure, and $\psi$
the so-called ``stress-angle'' denoting the angle from the horizontal
to the major principal stress. The components of the 2D stress tensor
are then
\begin{eqnarray}
  T_{xx} &=& -p(1+\sin\phi\cos 2\psi)
\\
 T_{yy}&=& -p(1-\sin\phi\cos 2\psi),
\\
T_{xy}=T_{yx}&=&-p\sin\phi\sin 2\psi
\end{eqnarray}
from which it can be seen that $p=-\tr \Tb/2$ and $\tan 2\psi=2
T_{12}/(T_{11}-T_{22})$. The convection-free 2D momentum balance law
$\del \cdot \Tb + \vec{F}_\textrm{body}=\vec{0}$ then reduces to the
two variable system of hyperbolic PDEs
\begin{eqnarray}
  (1+\sin\phi\cos2\psi)p_{x}-2p\sin\phi\sin2\psi\
  \psi_{x}  +\sin\phi\sin2\psi \ p_{y} +2p\sin\phi\cos2\psi\ \psi_{y}&=& F_\textrm{body}^x \\
  \sin\phi\sin2\psi\ p_{x}+2p\sin\phi\cos2\psi\ \psi_{x} +
  (1-\sin\phi\cos2\psi)p_{y} +2p\sin\phi\sin2\psi\ \psi_{y}&=&F_\textrm{body}^y.
\end{eqnarray}
A simple Mohr's circle analysis shows that the directions along which the yield
criterion is met, the slip-lines, form at the angles $\psi\pm\epsilon$ from the
horizontal where $\epsilon=\pi/4-\phi/2$.

\subsection{The stochastic flow rule}

To obtain flow from these stress equations, we will describe an
unorthodox, yet effective, ``stochastic flow rule'' (SFR) based on
postulating spot dynamics driven by local stress
imbalances~\cite{kamrin06}. Unlike classical flow rules in plasticity
which assume an ideal continuous medium, the SFR reflects the
discreteness and randomness inherent to a granular packing. In its
simplest embodiment, the SFR provides a unique way of defining the
spot drift $\Db_1$ and diffusivity $\Db_2$ in an arbitrary plane
strain environment. 

The SFR is based physically on the notion that spots move by sliding
on average along the (Mohr-Coulomb) slip-lines. However, there are two
crucial sources of randomness: The slip-lines themselves are blurred
by statistical fluctuations~\cite{ostoja05}, but, more fundamentally,
the spot must randomly decide between one of \emph{two} slip-lines
through each material point to slide in each step. Given the discrete
nature of the packing at the spot scale (several particle diameters),
it would be geometrically impossible for the spot to cause particle
displacements along both slip-lines at once.

In principle, the SFR is very general and could apply to any
limit-state stress field (not only for a granular material), but to
arrive at the simplest possible model for granular flow, with only one
parameter, we make several bold assumptions. The first is that of
isotropic spot diffusion $\Db_2=D_2\mathbf{1}$ which may be justified
to some extent by slip-line blurring.  To determine the magnitudes of
the drift and diffusion, we reason qualitatively. A spot can be viewed
as a local region of internal plastic failure, which perturbs the
contact stresses on neigboring material elements. Such a perturbation
can incite adjacent material to fail and when it does, the spot
effectively propagates to a new location. For a typical spot size
$L_s$, the distance of propagation should thus be roughly $L_s$ as
well.  Accordingly, we constrain all lengths referenced in the
definitions of the drift and diffusion, equations (\ref{eq:driftdef})
and (\ref{eq:diffdef}), to uphold this notion,
i.e. $\abs{\Db_1}=L_s/\Delta t$, and $D_2=L_s^2/2\Delta t$. All that
remains is to determine the drift direction $\ds=\Db_1/\abs{\Db_1}$,
which is where stresses enter the model.


We derive the spot drift from local stress imbalances upon
fluidization as follows (see \cite{kamrin06} for more details).
Define a material cell as the roughly diamond-shaped region
encompassed by two intersecting pairs of slip-lines, separated by
$L_s$. If a spot enters a cell, the value of the internal friction
along the cell boundaries should drop from $\mu$ to the kinetic value
$\mu_k$ causing a change in the shear tractions. This fluctuation may
generate a net force on the cell, which takes the simple form,
\begin{equation}\label{eq:fnet}
\Fb_\textrm{net}=\left(1-\mu_k/\mu\right)(\Fb_\textrm{body}-\cos^2\del p).
\end{equation}
This force pushes the material one way, and consequently a spot should
move the other. Since there is a slip-line field, however, a spot
cannot move in any arbitrary direction. Instead, the motion opposing
the net force must be projected onto slip-lines to obtain allowable
motions. By projecting $-\Fb_\textrm{net}$ onto the slip-lines and
averaging, we find that the spot drift direction is given by
\begin{eqnarray}\label{eq:bias}
\mathbf{\xi^{\pm}}&=&-(\Fb_\textrm{net}\cdot\mathbf{\hat{n}_{\psi\pm\epsilon}})
\mathbf{\hat{n}_{\psi\pm\epsilon}}
\\
\label{eq:bias2}
\ds&=&\frac{\mathbf{\xi^+}+\mathbf{\xi^-}}{\abs{\mathbf{\xi^+}+\mathbf{\xi^-}}}
\end{eqnarray}

With these assumptions, the SFR is specified enough for multiscale
simulations, alternating between continuum stress computation,
mesoscale spot random walks, and cooperative particle displacements.
Alternatively, as a continuum approximation, the mean flow profile can be
constructed by a two-step process:
\begin{enumerate}
\item{Calculate the steady spot concentration the drift-diffusion
  equation, which now has the simplified form
  \begin{equation}\label{eq:conc}
    \del\cdot(\ds\rho_s)=\frac{L_s}{2}\del^2\rho_s.
  \end{equation}}
\item{Compute the mean particle velocity by convolving the spot
    influence function with the reversed spot flux,
  \begin{equation}\label{eq:partvel}
    \ub=-\frac{L_s}{\Delta t}\int w(\rb,\rb')\left(\ds(\rb')\rho_s(\rb')
    -\frac{L_s}{2}\del\rho_s(\rb')\right)d\rb'.
  \end{equation}}
\end{enumerate}

As can be seen from above, the spot step duration $\Delta t$ has no
effect on the flow profile, aside from a simple rescaling of the
velocity. Together with the fact that equation (\ref{eq:conc}) is
homogeneous in $\rho_s$, we observe a consequence of the SFR common to
other rate-independent models: Any velocity field which solves the SFR
equations is unique only up to a multiplicative constant.

\section{Solving for Flow in Simple Geometries}
The SFR, with some theoretical exceptions as detailed in
\cite{kamrin06}, may apply as a general law in 2D plasticity. We will
focus on two canonical cases which represent two very different types of
granular flow: gravity-driven drainage from a silo and forced shear
flow in an annular Couette cell. We are not aware of any other model,
continuum or discrete, which can describe both of these cases, even
qualitatively, so this will be a stringent test of the SFR.  We
emphasize there will be no fitting parameters used. The parameters
$\phi$ and $L_s$ are independently measured material properties, and
their values are assumed not to depend on the flow environment.

\subsection{Silo}
\begin{figure}[t]
\begin{center}
(a)\epsfig{clip,width=2.5in, file=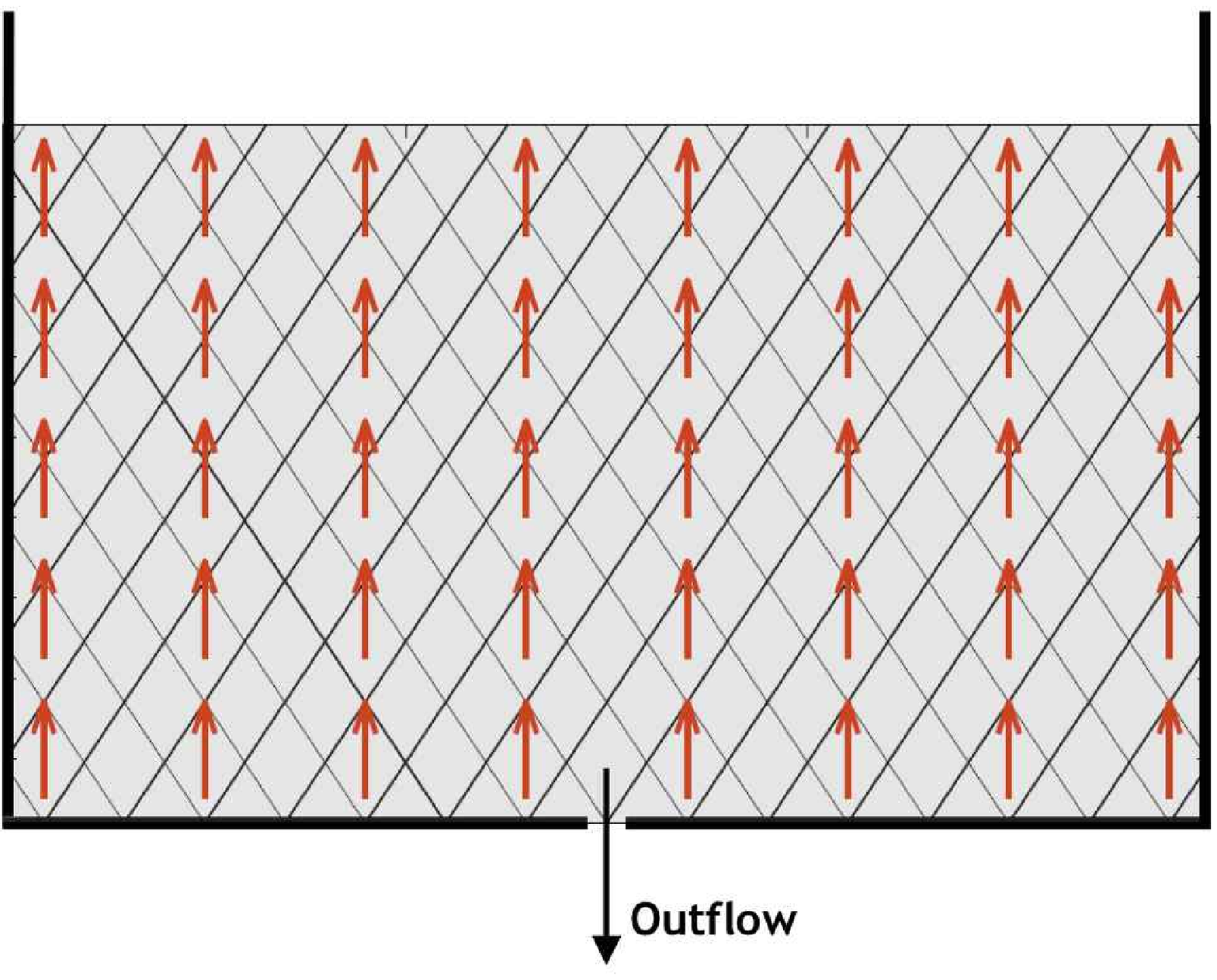} \ \ \ \ \ \ \ \ \ \ \ \ \ \ (b)\epsfig{clip,width=2.2in, file=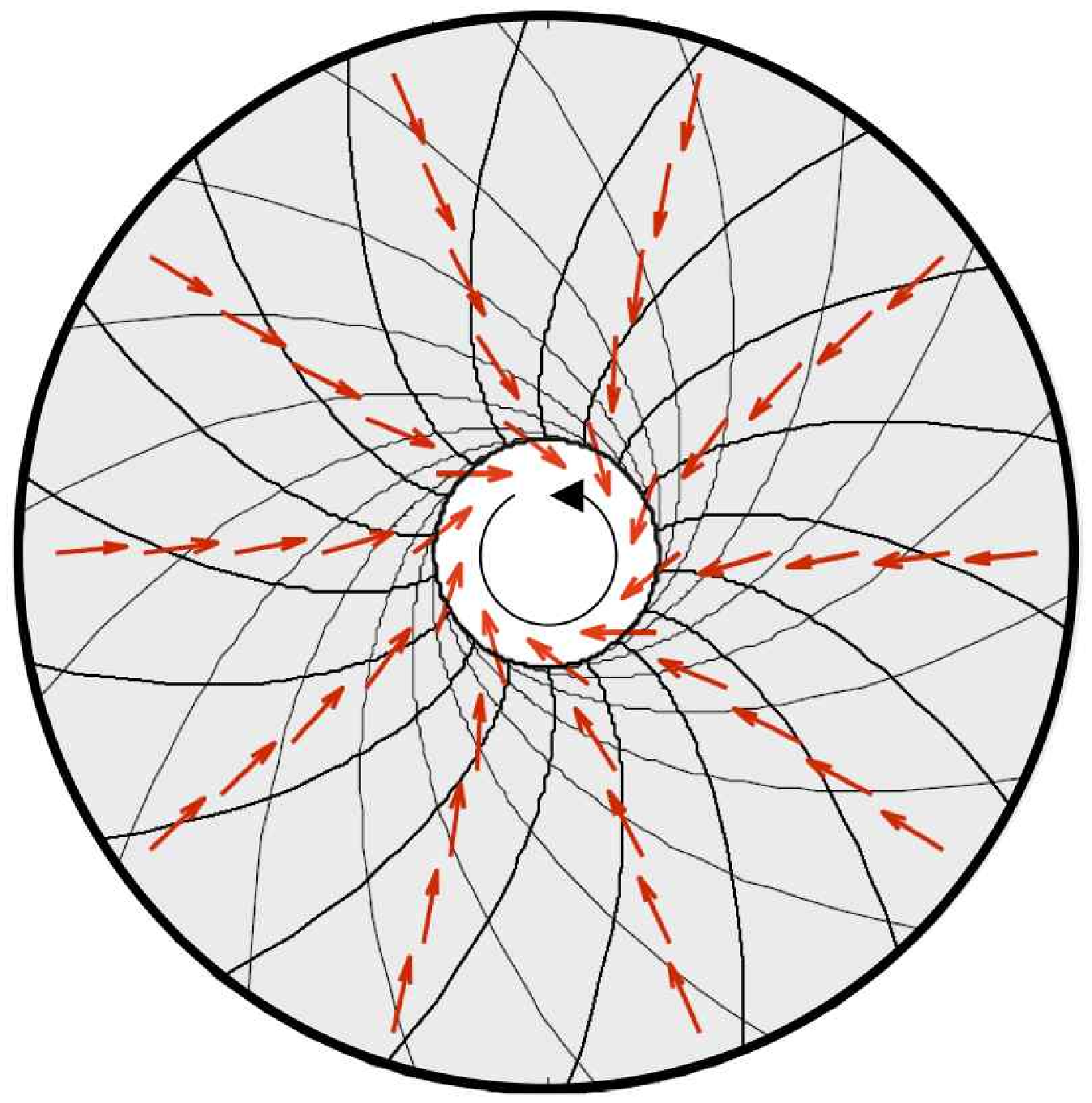}
\caption{Slip-lines fields (from MCP) and the spot drift field (from the SFR)
displayed for (a) a wide silo draining under gravity, and (b) shearing in an
annular Couette cell (no gravity).}\label{fig:drifts}
\end{center}
\end{figure}

For a wide, 2D silo with smooth walls and a flat bottom, the stress balance
equations can be solved analytically, giving
\begin{equation}
  \psi=\frac{\pi}{2}, \ \ \ \ \ p=\frac{f_g (z_{m}-z)}{1+\sin\phi}
\end{equation}
where $z$ is the distance from the silo bottom, $z_m$ is the distance from the
bottom to the free surface, and $f_g$ is the material's weight density.
Applying equation (\ref{eq:fnet}) gives $\Fb_\textrm{net}$ pointing uniformly
downward and thus $\ds=\mathbf{\hat{z}}$. To visualize the stress field (via
its slip-lines) simultaneously with the spot drift vector, see Figure
\ref{fig:drifts}(a). As such, the spot concentration solves
\begin{equation}
\frac{\p \rho_s}{\p z}=\frac{L_s}{2}\del^2\rho_s.
\end{equation}
For the narrowest possible silo opening, we have the boundary condition that
$\rho_s(x,0)=\delta(x)$. In line with our intuition on jammed states, this
implies the opening must be at least one spot width for any flow to occur.
Solving with a Fourier transform gives~\cite{kamrin06}
\begin{equation}
\rho_s \approx \frac{\exp(-x^2/4bz)}{\sqrt{4\pi b z}}
\end{equation}
for $b=L_s/2$ and consequently, by substituting $\rho_s$ into equation
(\ref{eq:unconv}), we have $\ub^*\cdot\mathbf{\hat{z}}\propto -\frac{\exp(-x^2/4bz)}{\sqrt{4\pi b z}}$.

Since the velocity gradient should not rapidly fluctuate in the bulk flow
region we are interested in, we utilize $\ub\approx\ub^*$. Diffusive spreading
of the downward velocity, as predicted here, is a well-documented phenomenon in
silo drainage \cite{nedderman79, samadani99, choi05, tuzun79, medina98a}.
Utilizing the $L_s$ dependence, the typical range of $b$ values is predicted to
be $1.5d$ to $2.5d$ which compares quite well to the results of documented
experiments on silo diffusion which, when accumulated, give a total $b$ range
of $1.3d$ to $3.5d$. To our knowledge, there is no other theory to 
predict $b$, even qualitatively, from basic mechanical principles, so
this should be viewed as a first success of the SFR. 

\subsection{Annular Couette cell}
A much stronger test of the SFR comes from its ability to adapt to a
completely different flow environment, without changing any
parameters.  In annular Couette flow, material fills the region
between two rough cylinders and is sheared by rotating the inner
cylinder while holding the outer stationary. To solve for the MCP
stresses, we first convert the stress balance equations to polar
coordinates $(r,\theta)$ and require that $p$ and $\psi$ obey radial
symmetry. This simplifies to
\begin{equation}
\frac{\p\psi^*}{\p r}=-\frac{\sin 2\psi^*}{r(\cos 2\psi^* +\sin\phi)}, \  \ \ \ \ \frac{\p\eta}{\p r}=-\frac{2\sin \phi}{r(\cos 2\psi^*+\sin \phi)}
\end{equation}
where $\eta=\log p$ and $\psi^*=\psi+\frac{\pi}{2}-\theta$. We solve
these equations numerically using fully rough inner wall boundary
conditions $\psi^*(r_w)=\frac{\pi}{2}-\epsilon$ and any arbitrary
value for $\eta(r_w)$.  We find that $p$ decreases with distance from
the inner wall, implying $\Fb_\textrm{net}$ points radially
outward. Applying equations (\ref{eq:bias}) and (\ref{eq:bias2}) gives
a drift field that points inward, but gradually opposes the motion of
the inner wall.

Below, we solve for $\ub^*$ directly by enforcing that the flow point
along $\vec{\hat{\theta}}$. Since the material near the wall shears
rapidly, we cannot assert $\ub\approx\ub^*$ and must solve the full
convolution, equation (\ref{eq:ub}). For simplicity, we assume
circular spots $w(\rb,\rb')\propto H(\frac{L_s}{2}-\abs{\rb-\rb'})$
where $H$ is a step function. (Velocity correlations measurements in
silo experiments suggest that the spot shape may be anisotropic due to
gravity, or more generally, local stresses, but this is not a major
effect~\cite{kamrin06}.)  To properly describe particle motion near
the wall, we must decide how spots interact with walls so that the
convolution integral can be carried out all the way to the inner
wall. A simple claim is that $\ub^*$ must equal the wall velocity
wherever spots overlap the wall---a no-slip condition of sorts. After
convolution, the net effect of this assertion is that $\ub$ is
slightly flatter than $\ub^*$ close to the inner wall.  We shall refer
to this assumption about wall interactions as ``wall hypothesis
1''. We note that this is slightly different than the hypothesis used
in \cite{kamrin06} which claims the region within the walls can be
viewed as a bath of non-diffusive spots which cause particles to move
in a manner that mimics the rigid wall motion. We shall refer to this
as ``wall hypothesis 2''.  Though our theory is flexible in terms of
near wall specificities, the bulk behavior, regardless of which
hypothesis is selected, is always the same.

\section{Comparing SFR Predictions to Discrete Element Simulations}
\subsection{Simulation Methods}\label{sec:model}
To test the above models, we carried out DEM simulations using a
modified version of the Large-scale Atomic/Molecular Massively
Parallel Simulator (LAMMPS) developed at Sandia National Laboratories
\cite{lammps}. Following a similar approach to previous work by our
group \cite{rycroft06a,reactorsim} and others
\cite{silbert01,landry03} we considered simulations of spherical
particles interacting using a modified version of the model developed
by Cundall and Strack \cite{cundall79} to model cohesionless
particulates. Monodisperse spheres with diameter $d$ interact
according to Hertzian, history dependent contact forces. For a
distance $\vec{r}$ between a particle and its neighbor, when the
particles are in compression, so that $\delta=d-\abs{\vec{r}}>0$, then
the two particles experience a force $\vec{F}=\vec{F}_n+\vec{F}_t$,
where the normal and tangential components are given by
\begin{equation}
  \vec{F}_n=\sqrt{\delta/d} \left(k_n\delta\vec{n} - \frac{\gamma_n\vec{v}_n}{2}\right), \ \ \ \ \vec{F}_t=\sqrt{\delta/d} \left(-k_t\Delta\vec{s}_t -\frac{\gamma_t\vec{v}_t}{2}\right).
\end{equation}
Here, $\vec{n}=\vec{r}/\abs{\vec{r}}$. $\vec{v}_n$ and $\vec{v}_t$ are the
normal and tangential components of the relative surface velocity, and
$k_{n,t}$ and $\gamma_{n,t}$ are the elastic and viscoelastic constants,
respectively. $\Delta\vec{s}_t$ is the elastic tangential displacement between
spheres, obtained by integrating tangential relative velocities during elastic
deformation for the lifetime of the contact, and is truncated as necessary to
satisfy a local Coulomb yield criterion $\abs{\vec{F}_t}\le \mu_c
\abs{\vec{F}_n}$. Particle-wall interactions are treated identically, though
the particle-wall friction coefficient $\mu_w$ is set independently.

For the current simulations we set $k_t=\frac{2}{7}k_n$, and choose
$k_n=2\times10^5 mg/d$. This is significantly less than would be
realistic for many hard granular materials, such as glass beads, where we
expect $k_n > 10^{10} mg/d$, but such a spring constant would be prohibitively
computationally expensive, as the time step scales as $\delta t \propto
k_n^{-1/2}$ for collisions to be modeled effectively. Previous simulations have
shown that increasing $k_n$ does not significantly alter physical results
\cite{landry03}. We use a time step of $\delta t=1.0\times10^{-4}\tau$ and
damping coefficients $\gamma_n=50\tau^{-1}$ and $\gamma_t=0.0$. All
measurements are expressed in terms of $d$ and $\tau$. We make use of both
cartesian coordinates $(x,y,z)$ and cylindrical coordinates $(r,\theta,z)$
where $r^2=x^2+y^2$; gravity is set to $1 d\tau^{-2}$ and points in the
negative $z$ direction.

The simulations were carried out on MIT's Applied Mathematics Computational
Laboratory, a Beowulf cluster consisting of 16 nodes each with a dual
processor. During the simulations, snapshots of all particle positions were
saved every $2\tau$, corresponding to 20000 integration timesteps. The LAMMPS
code is written in C++ and can be run on any number of processors, by
decomposing the computational domain into cuboidal subdomains of equal size.
Interactions between particles in neighboring domains are handled using message
passing. For problems where the particles are split evenly between the
subdomains, the LAMMPS code scales very well, and doubling the number of
processors can frequently result in almost a doubling of speed. However, the
Couette geometry causes some subdomains to have more particles than others,
leading to poor load-balancing. Most of the simulations were therefore run
on four processors, with the domains split by the planes $x=0$ and $y=0$ to
achieve optimal load-balancing; these simulations took approximately four weeks
to complete.


\subsection{The silo geometry}
\begin{figure}[t]
\begin{center}
(a)\epsfig{clip,width=4.33in, file=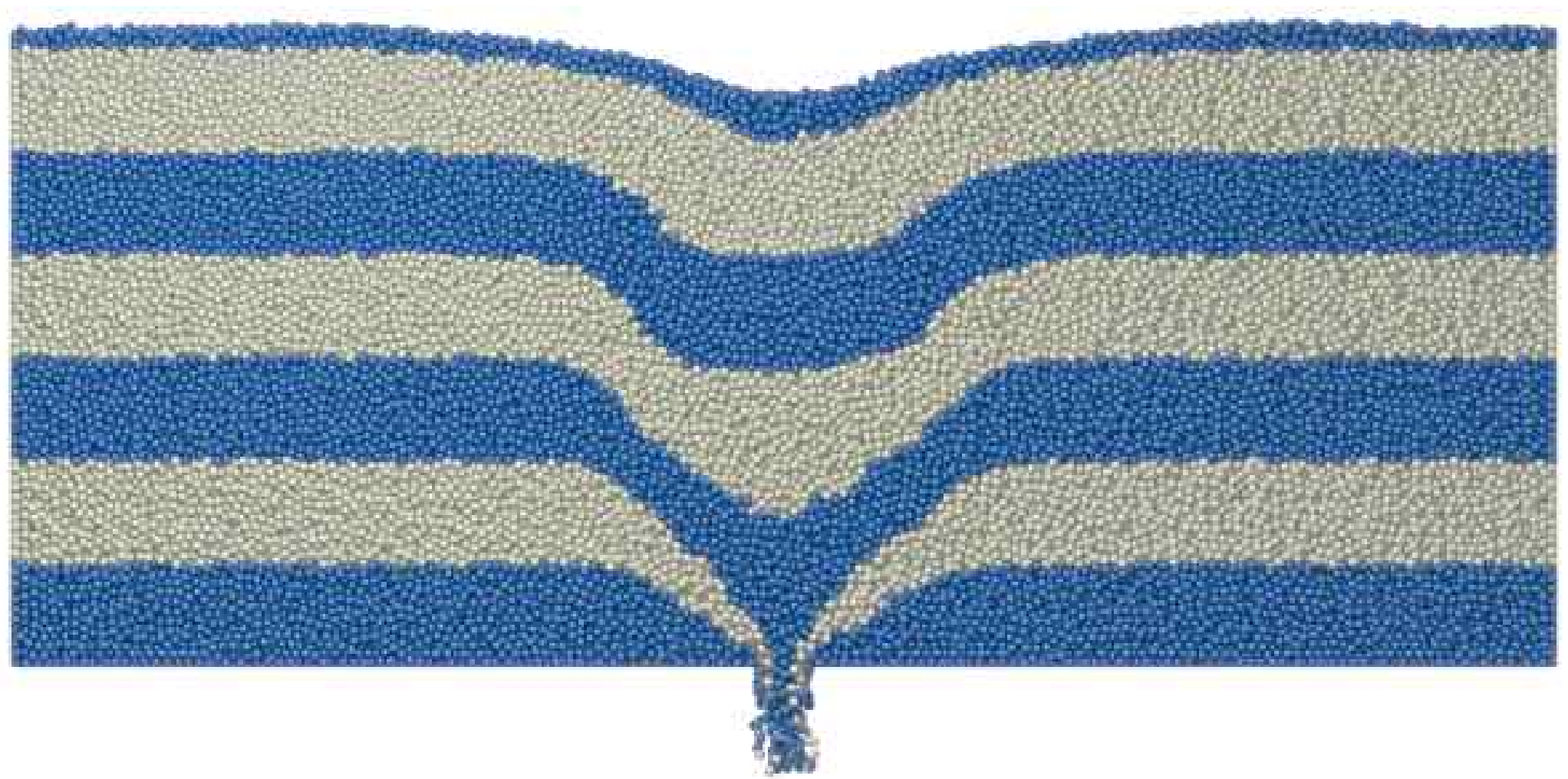}
  \\
(b) \epsfig{clip,width=3in, file=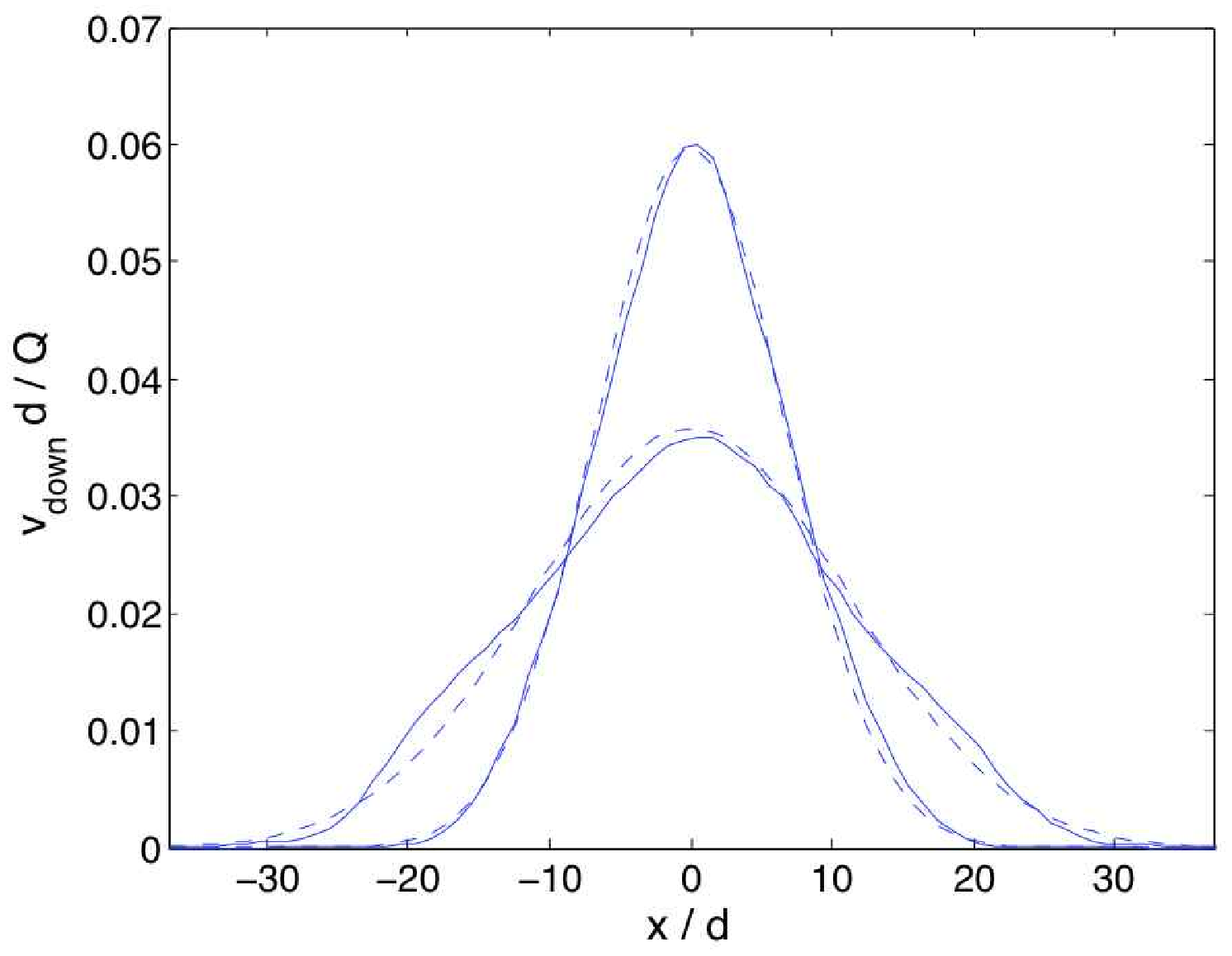}  (c) \epsfig{clip,width=3in, file=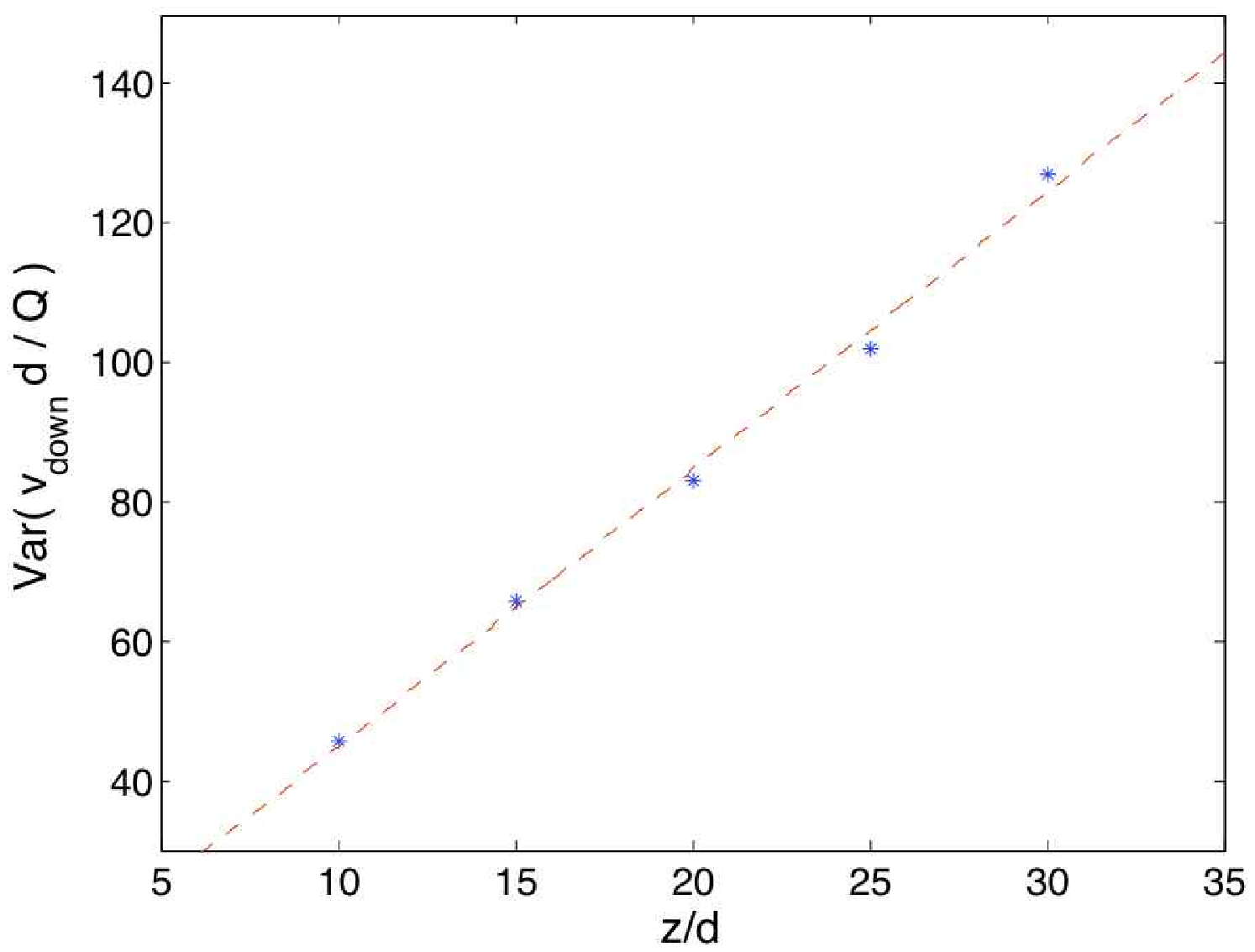}
\caption{(a) A typical snapshot of the silo system during drainage,
  taken at $t=60\tau$. The colored bands are initially spaced $10d$
  apart, and highlight the deformations that occur during flow (b)
  Mean downward velocity profiles at two different heights in a wide
  silo. DEM results (solid lines) are plotted against the SFR
  predictions (dashed lines) at heights of $z=10d, 30d$; (c) Mean
  square width of the downward velocity across horizontal
  cross-sections in the DEM (points), compared to the SFR (dashed
  line)} \label{fig:silo_vprofile}
\end{center}
\end{figure}

We considered a quasi-2D silo with plane walls at $x=\pm 75d, z=0$
with friction coefficient $\mu_w=0.5$, and made the $y$ direction
periodic with width $8d$.  To generate an initial packing, $90,000$
spherical particles with contact friction coefficient $\mu_c=0.5$ were poured in
from a height of $z=130d$ at a rate of $378\tau^{-1}$.  After all
particles are poured in at $t=238\tau$, the simulation is run for an
additional $112\tau$ in order for the particles to settle. After this
process has taken place, the particles in the silo come to a height of
approximately $z=62.2d$. To initiate drainage, an orifice in the base
of the container is opened up over the range $-3d<x<3d$, and the
particles are allowed to fall out under gravity; Figure
\ref{fig:silo_vprofile}(a) shows a typical simulation snapshot during
drainage.

As noted in the previous section, snapshots of the particle positions are saved
every $2\tau$. We collected $282$ snapshots, and made use of this information
to construct velocity cross sections. A particle with coordinates $\vec{x}_n$
at the $n$th timestep and $\vec{x}_{n+1}$ at the $(n+1)$th timestep makes a
velocity contribution of $(\vec{x}_{n+1}-\vec{x}_n)/\Delta t$ at the point with
coordinates $(\vec{x}_n+\vec{x}_{n+1})/2$. This data can then be appropriately
binned to create a velocity profile; we considered bins of size $d$ in the $x$
direction, and created velocity profiles for different vertical slices
$\abs{z-z_s}<d/2$.

Since the SFR makes predictions about the velocity profile during
steady flow, we choose a time interval $t_1<t<t_2$ over which the
velocity field is approximately constant. Choosing this interval
requires some care, since if $t_1$ is too small then initial
transients in the velocity profile can have an effect, and if $t_2$ is
too large, then the free surface can have an influence.  For the
results reported here, we chose $t_1=120\tau$ and $t_2=200\tau$.

Figure \ref{fig:silo_vprofile}(b) compares the SFR predictions for
this environment to the DEM simulation. The displayed simulation data
uses a particle contact friction of $\mu_c=0.3$. Since the typical range
of $L_s$ from velocity correlations in simulations~\cite{rycroft06a}
and experiments~\cite{kamrin06} is $3d$ to $5d$, we choose $L_s=4d$ to
generate the approximate SFR solution. We emphasize that this
parameter is not fitted. In this geometry, the slip-lines are
symmetric about the drift direction causing both $\Db_1$ and $D_2$ to
become independent of the internal friction. In prior simulations we
have found that that particle contact friction has some effect on the
flow~\cite{reactorsim}, and analogously the internal friction should
play some role in the determination of $b$. Here, the loss of friction
dependence comes from our simplification that $\Db_2$ is isotropic. A
less simple but more precise definition for $\Db_2$ would
anisotropically skew the spot diffusion tensor as a function of
internal friction: the slip-lines, which we model as the directions
along which a spot can move (roughly), intersect at a wider angle as
internal friction is increased.

Even so, our simple model captures many of the features of the flow
and accurately portrays the dominant behavior. The downward velocity,
especially at $z=10d$, strongly matches the predicted
Gaussian. Perhaps a more global demonstration of the underlying
stochastic behavior in the SFR is evident in Figure
\ref{fig:silo_vprofile}(c), where a linear relationship can be seen
between the mean square width of $v_z$ and the height, indicating that
the system variables are undergoing a type of diffusive scaling. The
SFR solution also predicts this linear relationship and, in
particular, that the slope should equal $2b=L_s$.  The agreement shown
in Figure 7(b) for such a typical $L_s$ value is a strong indicator
that the role of the correlation length in the flow has been properly
accounted for in the SFR.

The diffusive velocity profile near the orifice of a wide silo has
also been observed in a number of
experiments~\cite{mullins74,tuzun79,medina98b,samadani99} and DEM
simulations~\cite{rycroft06a,reactorsim} and was first predicted
qualitatively by kinematic
models~\cite{lit58,mullins72,nedderman79}. Such models, however, lack
a plausible microscopic basis~\cite{bazant06} or any connection with
mechanics and cannot be applied to any other geometry, other than the
wide silo.

\subsection{The Couette geometry}
For the Couette geometry, we considered five different interparticle friction
coefficients, $\mu_c=0.1, 0.3, 0.5, 0.7, 0.9$, and for each value an initial
packing was generated using a process similar to that for the silo. We consider
a large cylindrical container with a side wall at $r=64d$ with friction
coefficient $\mu_w=\mu_c$, and a base at $z=0$ with friction coefficient
$\mu_w=0$. For each simulation, 160000 particles are poured in from a height of
$z=48d$ at a rate of $4848\tau^{-1}$. After all particles are introduced at
$t=33\tau$, the simulation is run for an additional time period of $322\tau$ to
allow the particles to settle. After this process has taken place, the initial
packings are approximately $11.5d$ thick. Packings with $\mu_c=0.9$ are
approximately $2\%$ thicker than those with $\mu_c=0.1$.


\begin{figure}
  \begin{center}
  (a) \epsfig{clip,width=3in, file=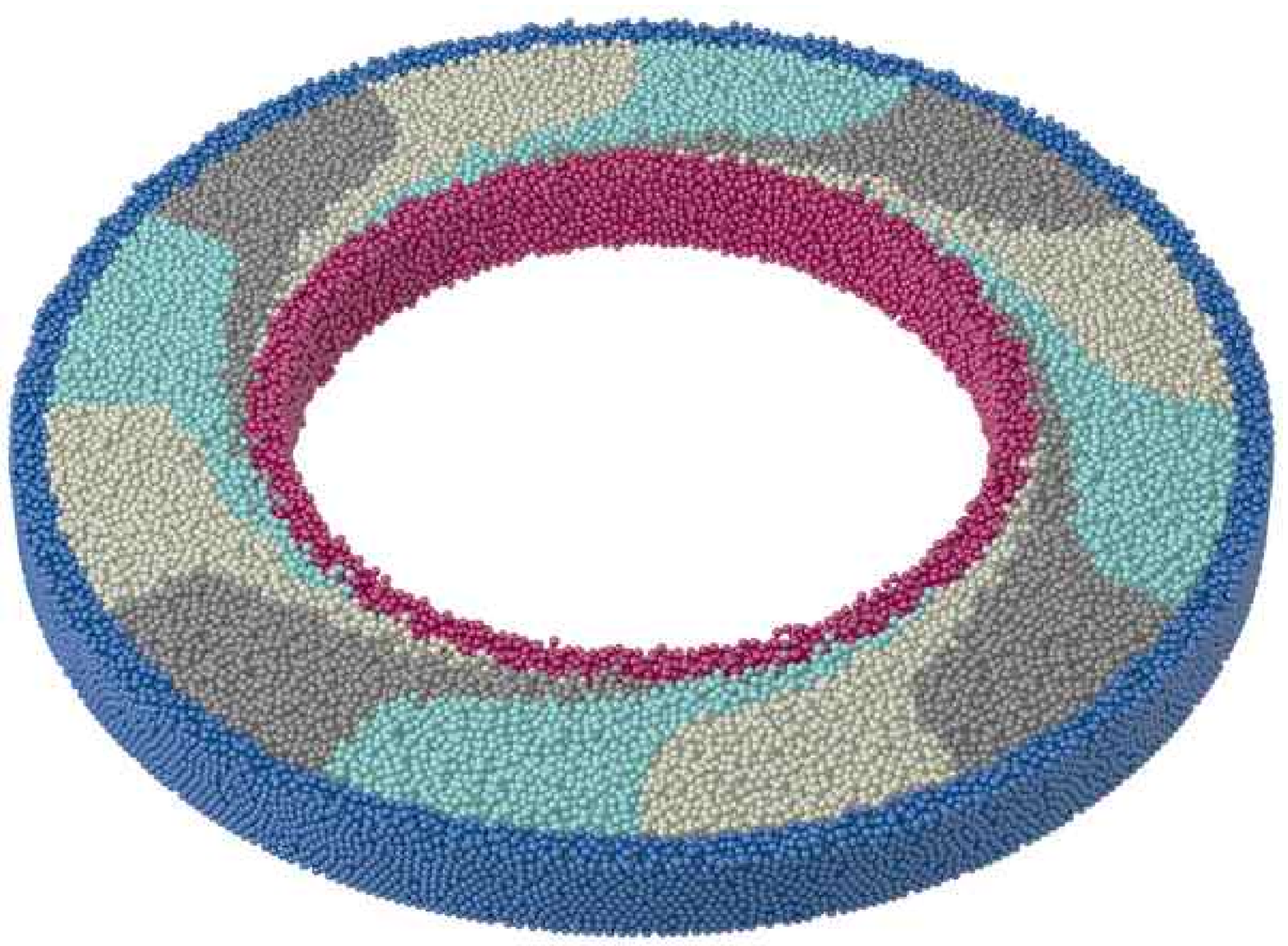}  \ (b)\epsfig{clip,width=3in, file=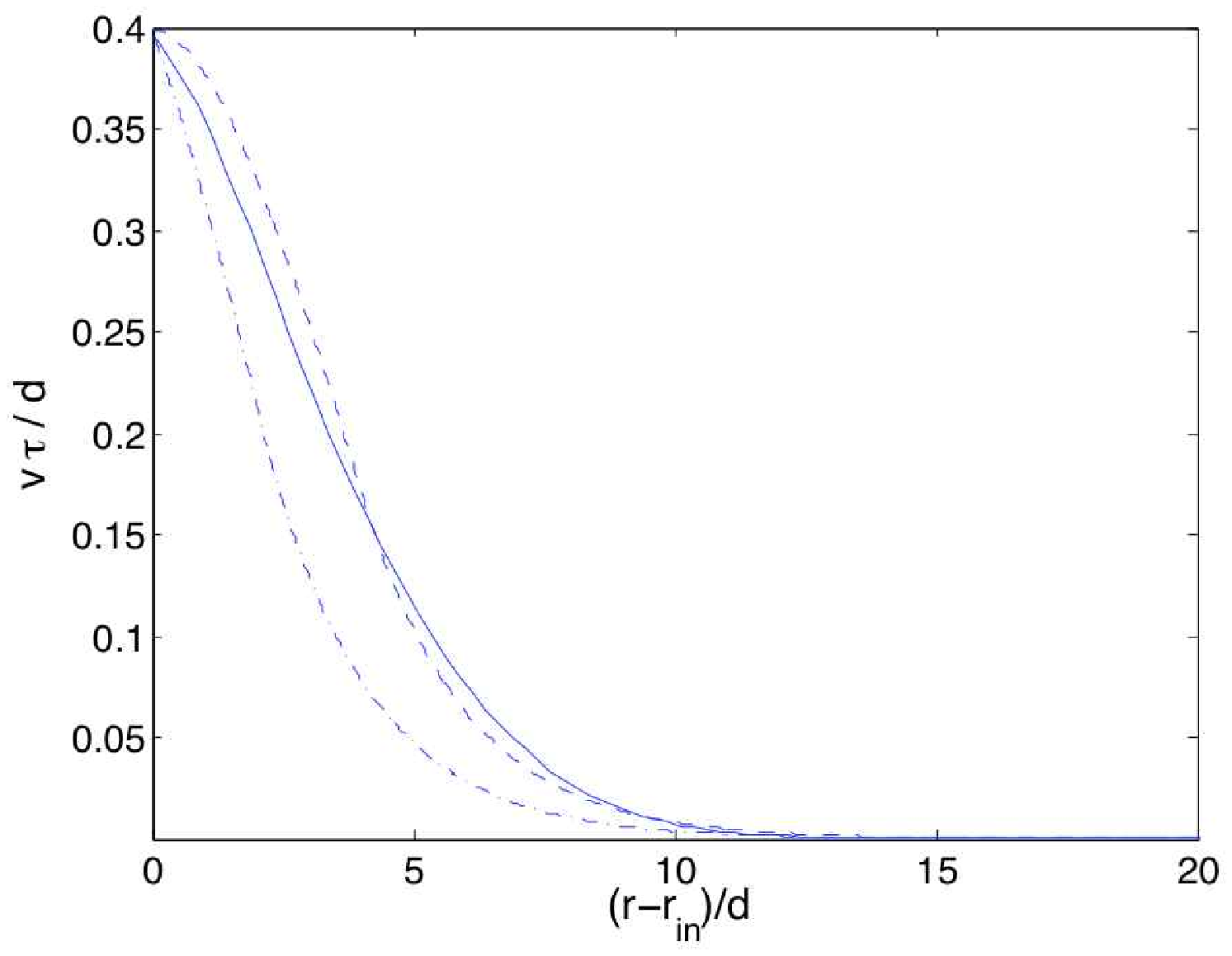}
  \caption{(a) A typical snapshot of the system during shearing based
    on the simulation with $\rin=40d$, $\omega=0.01\tau^{-1}$, and
    $\mu_c=0.5$. Dark blue particles (with $r>\rout$) are frozen
    during the simulation, while dark red particles (with $r<\rin$)
    are rotated with angular velocity $\omega$. The particles between
    $\rin$ and $\rout$ undergo shearing. The colored bands of grey,
    cream, and cyan were initially radial, and in this snapshot, after
    fifty frames, the deformations can be clearly seen. (b) SFR
    solutions using wall hypothesis 1 (dashed line) and wall
    hypothesis 2 (dot-dashed line) in the annular Couette geometry
    ($\rin=40d, \rout=60d$) are compared to the simulation results
    (solid line). Both use the same parameters that were used in the
    silo geometry ($L_s=4d, \mu_c=0.3$) and the bead internal friction
    angle is set at the typical value $25^o$.}\label{fig:annular}
\end{center}
\end{figure}
The shearing simulations take place in an annulus between two radii, $\rin$ and
$\rout$. A rough outer wall is created by freezing all particles which have
$r>\rout$; any forces or torques that these particles experience are zeroed
during each integration step. Similarly, all particles between $\rin-4d$ and
$\rin$ are forced to rotate with a constant angular velocity $\omega$ around
$r=0$. Particles with $r<\rin-4d$ are deleted from the simulation and an extra
cylindrical wall with friction coefficient $\mu_w=\mu_c$ is introduced at
$r=\rin-4d$ to prevent stray surface particles from falling out of the shearing
region. A typical run is shown in Figure \ref{fig:annular}(a).

From the five initial packings, we carried out eight different shear cell
simulations to investigate the effects of friction, angular velocity, and inner
wall radius. To investigate the effect of friction, five runs were carried out
with $\omega=0.01\tau^{-1}$ and $\rin=40d$, for $\mu_c=0.1, 0.3, 0.5, 0.7, 0.9$.
To examine the effect of the inner wall radius, an additional run with
$\rin=30d$ and $\rout=50d$ was carried out, with $\mu_c=0.3$ and
$\omega=0.01\tau^{-1}$ kept constant. To look at the effect of angular
velocity, a further two runs with $\omega=0.05\tau^{-1}$ and
$\omega=0.2\tau^{-1}$ were carried out, with $\mu_c=0.5$ and $\rin=40d$ kept
constant. For each simulation, we collected $561$ snapshots. For the runs where
$\rin=40d$, approximately $108350$ particles were simulated, corresponding to
$2.0$Gb of data. For the run with $\rin=30d$, $88657$ particles were simulated,
corresponding to $1.7$Gb of data.

The simulation results show a good empirical agreement with previous
experimental work on shear cells \cite{losert00, midi04, latzel03,
  bocquet01, mueth00}. In all cases, we see an angular velocity
profile that falls off exponentially from the inner cylinder, with a
half-width on the order of several particle diameters. Near the inner
wall, the flow deviates from exponential, which is an effect seen in
some prior studies but is more dramatic here.  Following methods
similar to those used in silo simulation, we used the snapshots to
construct an angular velocity profile. We used bins of size $d/2$
in the radial direction,
and we looked at velocity profiles in different vertical slices
$\zlow<z<\zhigh$.

Since the simulation geometry is rotationally symmetric, our angular velocity
profile can most generally be a function of $r$, $z$ and $t$. Ideally, we hope
that $\omega$ is primarily a function of $r$, with only a very weak dependence
on $z$ and $t$, but we began by studying the effects of these other variables.
To determine the dependence on time, the velocity profiles were calculated over
many different time intervals. As would be expected, the simulation had to be
run for small amount of time before the velocity profile would form; this
happens on a time scale of roughly $50\tau$, and the results suggest a longer
time is needed for the cases with low friction. However, the data also shows
time-dependent effect happening on a longer scale: as the shearing takes place,
there is a small but consistent migration of particles away from the rotating
wall, which has a small effect on the velocity profile. This effect does
eventually appear to saturate, but because of this, we chose to discard the
simulation data for $t<500\tau$ and calculate velocity profiles based on the
time window $500\tau<t<1100\tau$.

To investigate the angular velocity dependence on the height, we calculated the
velocity profiles in five different slices $z_h<z<z_h+d$ for $z_h=d, 3d, 5d,
7d, 9d$. Near the inner rotating wall, the velocity profiles show very little
dependence on height. However, in the slow-moving region close to the fixed
outer wall, large differences can be seen, with particles in the lowest slice
moving approximately $30\%$ slower than those in the central slice, and those
in the top slice moving approximate $30\%$ faster. The three central slices
show differences of at most $10\%$, and we therefore chose to use the range
$3d<z<8d$.

The SFR treats the correlation length as a material property independent of the
flow geometry or other state variables. To see how well this notion is upheld,
we solve the SFR in the annular Couette geometry using the same correlation
length ($L_s=4d$) that was used in the displayed silo prediction, Figure
\ref{fig:silo_vprofile}. It is then compared to a simulation of annular flow
which uses the same grain properties ($\mu_c=0.3$).

Results from Figure \ref{fig:annular}(b) show that regardless of the wall
hypothesis, the SFR prediction captures the same qualitative features
of the simulation.  The SFR and simulation both predict a flatter
range near the inner wall, followed by exponential decay. Near the
inner wall, it does appear that wall hypothesis 1 (``no slip
condition'' for spots) gives a closer match to the simulation.

\begin{figure}
  \centering
  \epsfig{clip,width=2.8in, file=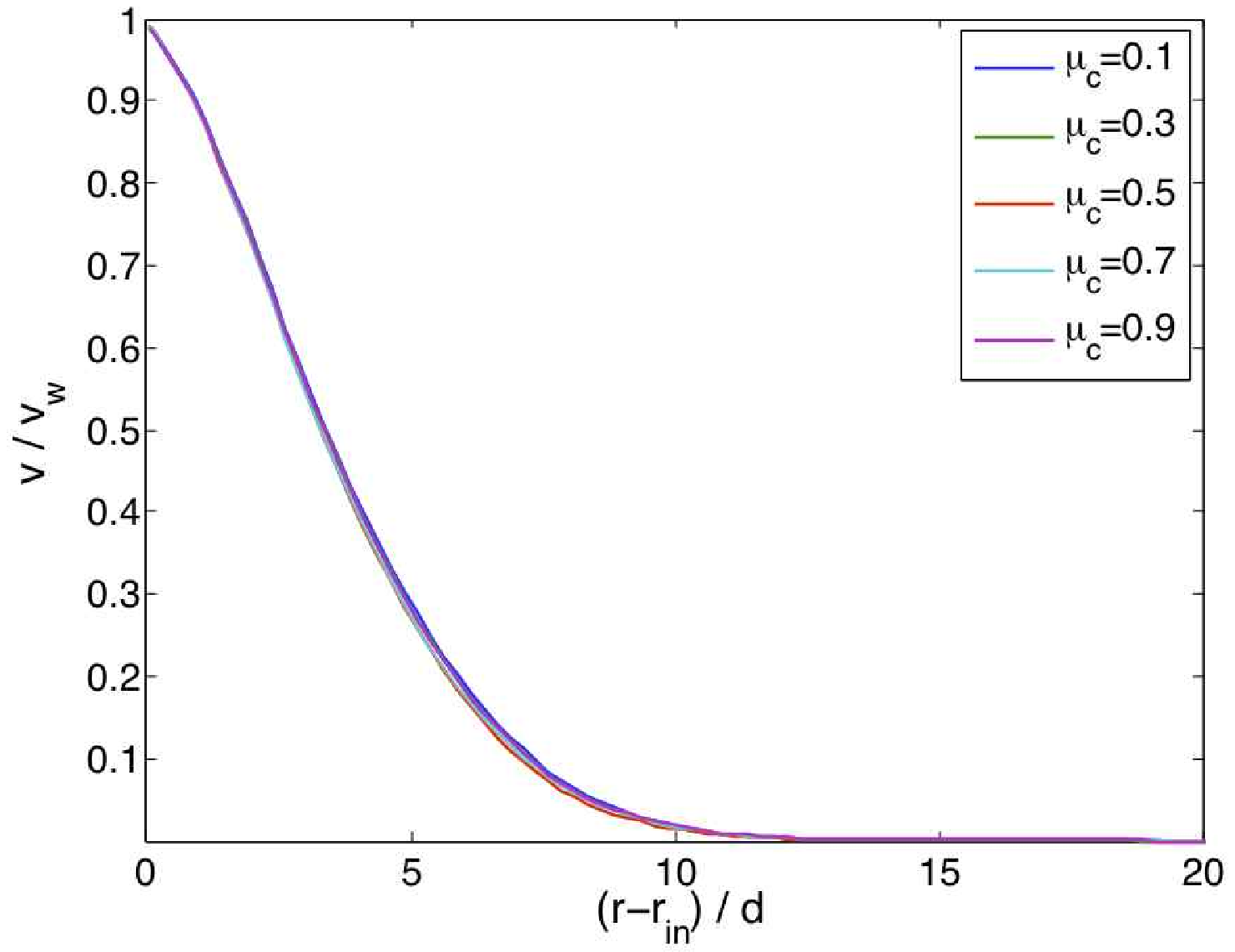} \ \ \epsfig{clip,width=2.8in, file=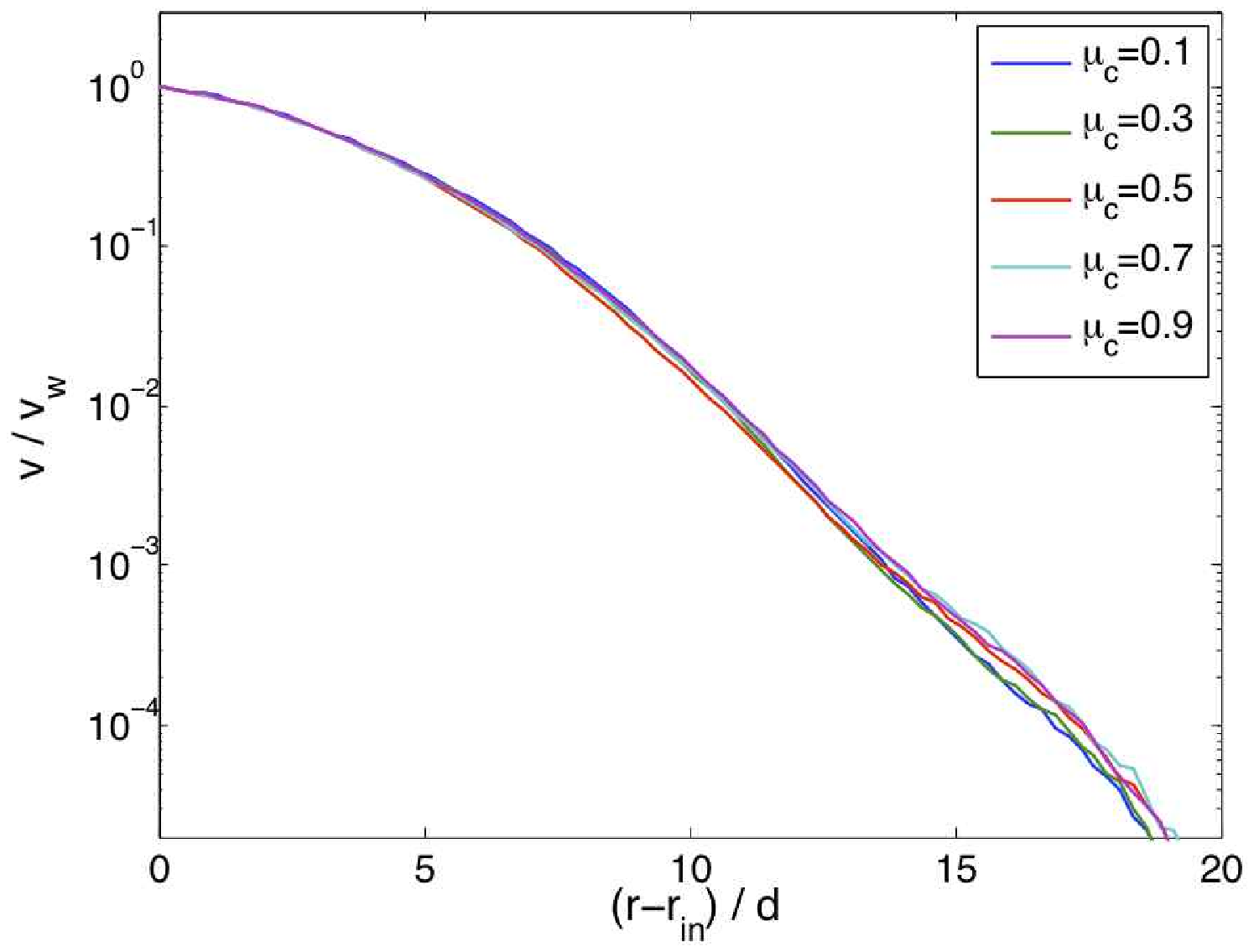}
  \caption{Velocity profiles for five different values of $\mu_c$, where
  $\rin=40d$ and $\omega=0.01\tau^{-1}$.}
  \label{fig:profiles_mu}
\end{figure}

The SFR, when applied to the annular flow geometry, does predict a
slight dependence of the flow on the internal friction. We emphasize
that internal friction is not the same quantity as particle contact
friction $\mu_c$, though we believe if the contact friction is
increased, inevitably, the internal friction must be as
well. Spherical grains almost always have internal friction angles in
the range $\phi=20^\circ$ to $\phi=30^\circ$, so to represent this
range as best as possible, we simulated flows varying the particle
contact friction from $\mu_c=0.1$ to $\mu_c=0.9$. Figure
\ref{fig:profiles_mu} displays velocity profiles for
five different values of friction. In the semi-log format, profiles
appear almost linear over the range $45d<r<58d$ indicating a very good
fit to an exponential model of velocity.

Since the curves in the figure are very close, and exhibit some experimental
noise, it is difficult to discern any small differences in the widths of the
velocity profiles. However, table \ref{tab:halfwidths} shows the results of
applying linear regression to extract a half-width for each velocity profile.
We see differences on the order of $5\%$, roughly in line with the SFR. More
importantly, the trend of increasing flow width with increasing friction is
seen in both.

\begin{figure}\label{radius}
\begin{center}
\epsfig{clip,width=3.2in, height=2.6in, file=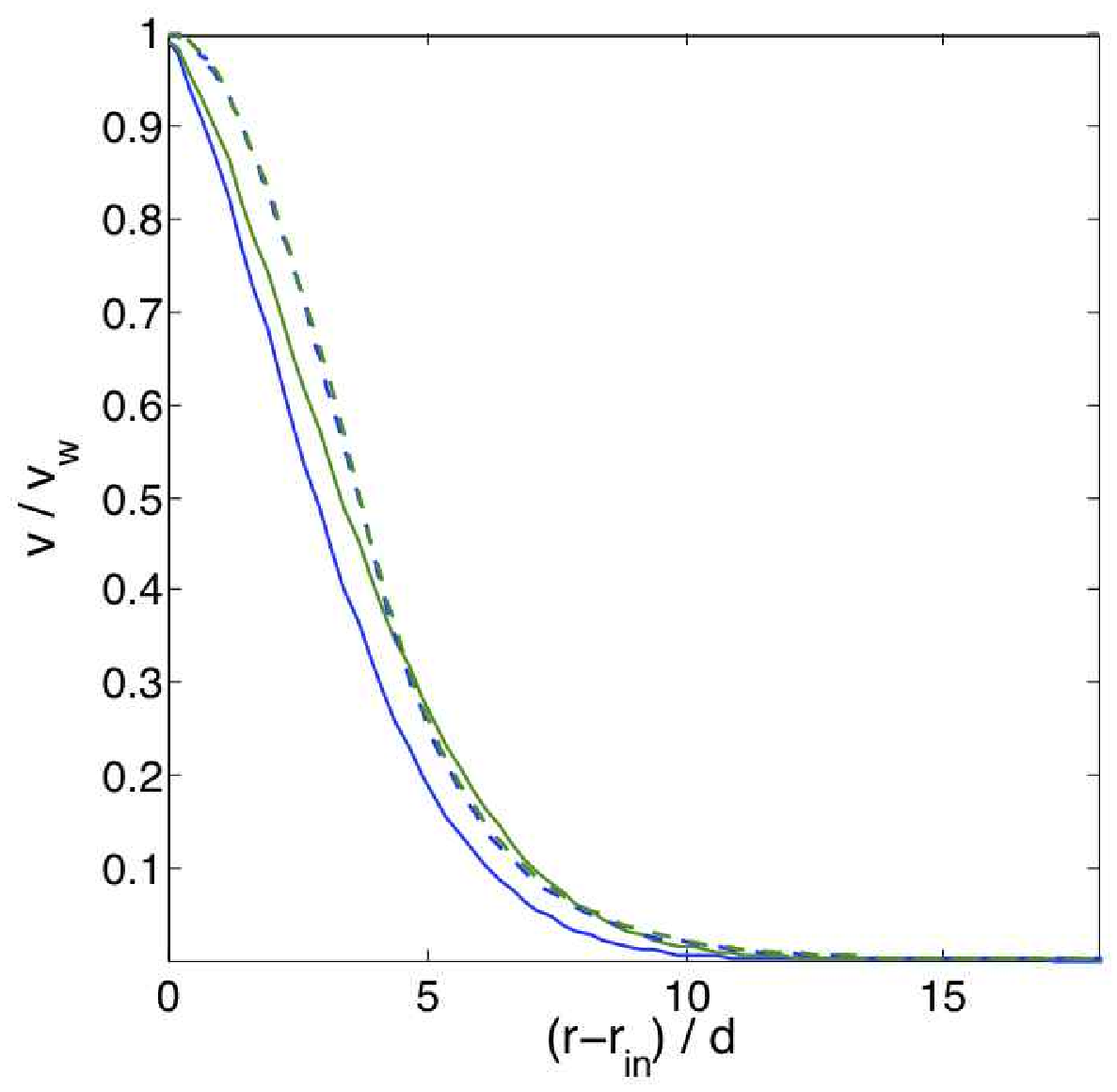}  \epsfig{clip,width=3.2in, file=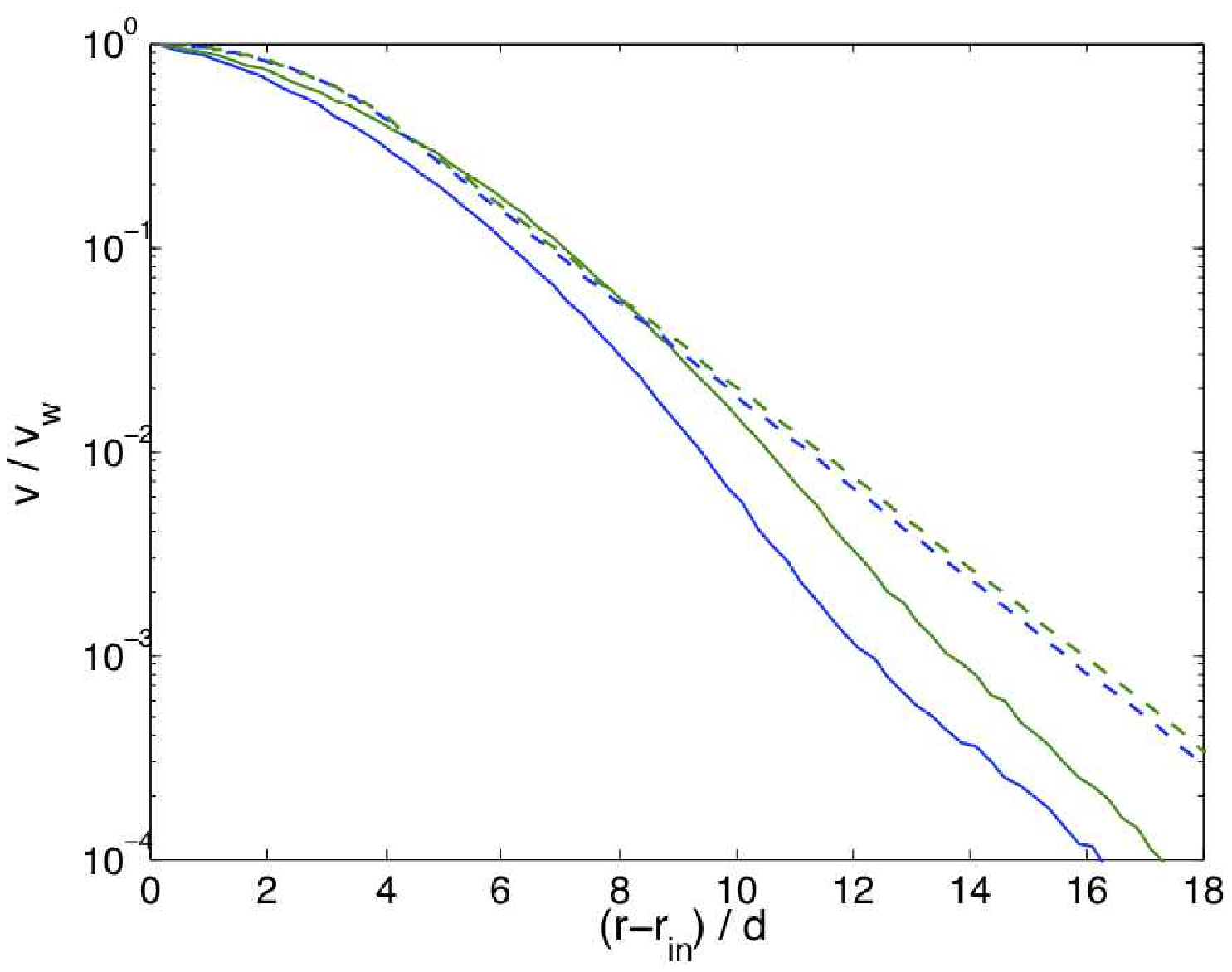}
\caption{DEM simulation results (solid lines) for the flow profile
  when the inner wall radius is 30d (blue) and 40d (green) compared to
  SFR predictions (dashed lines) which use the same color scheme.  For
  solidarity with the silo flow predictions, the SFR solutions use
  $L_s=4d$. }\label{annradius}
\end{center}
\end{figure}

When the inner wall radius is decreased, Figure \ref{annradius}
indicates that the shear band shrinks but the decay behavior in the
tail changes only minimally.  The SFR predicts this qualitative trend
as well, but significantly underestimates the size of the shear-band
decrease.

\begin{figure}
  \centering
  \epsfig{clip,file=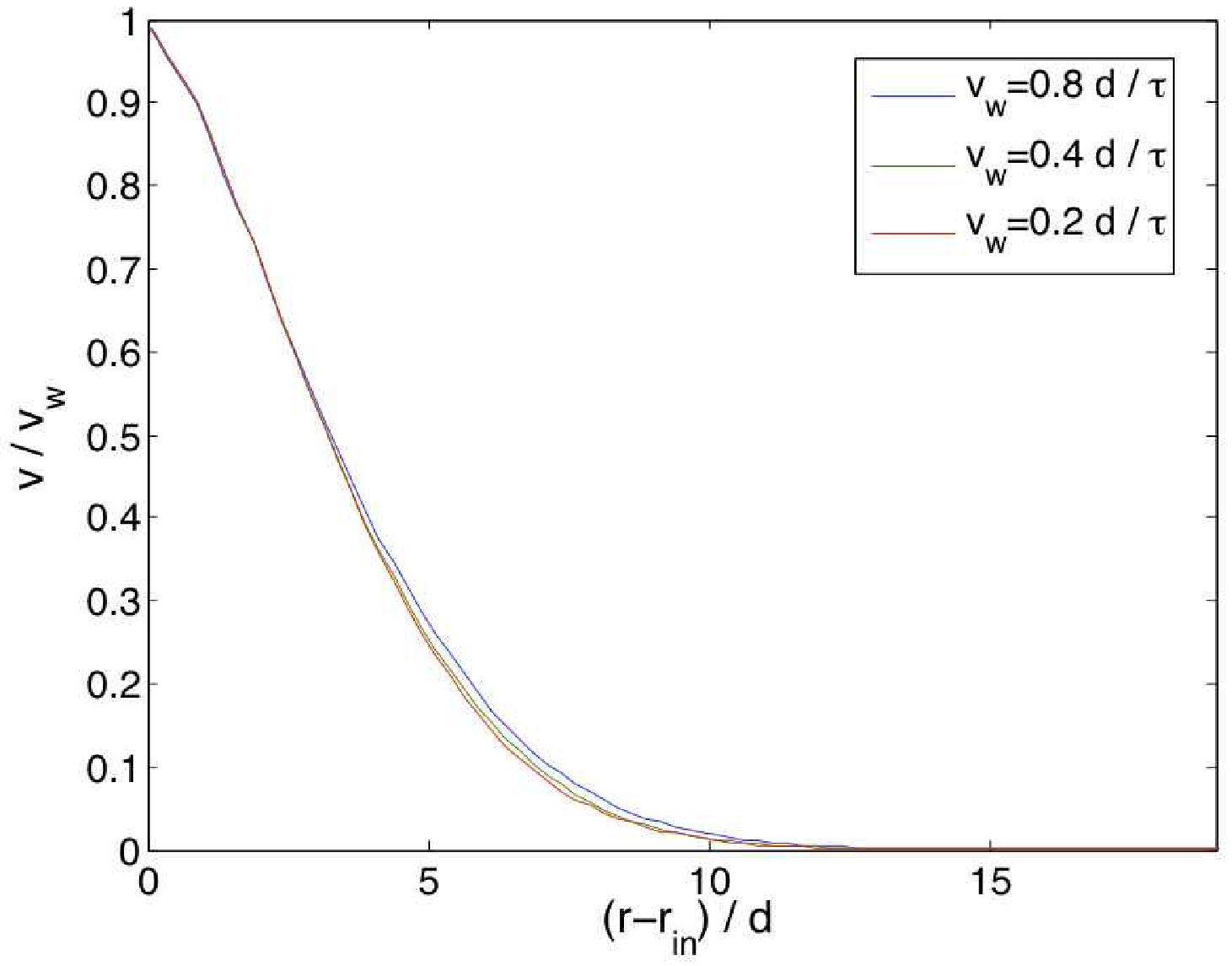, width=3in} \ \ \ \epsfig{clip,file=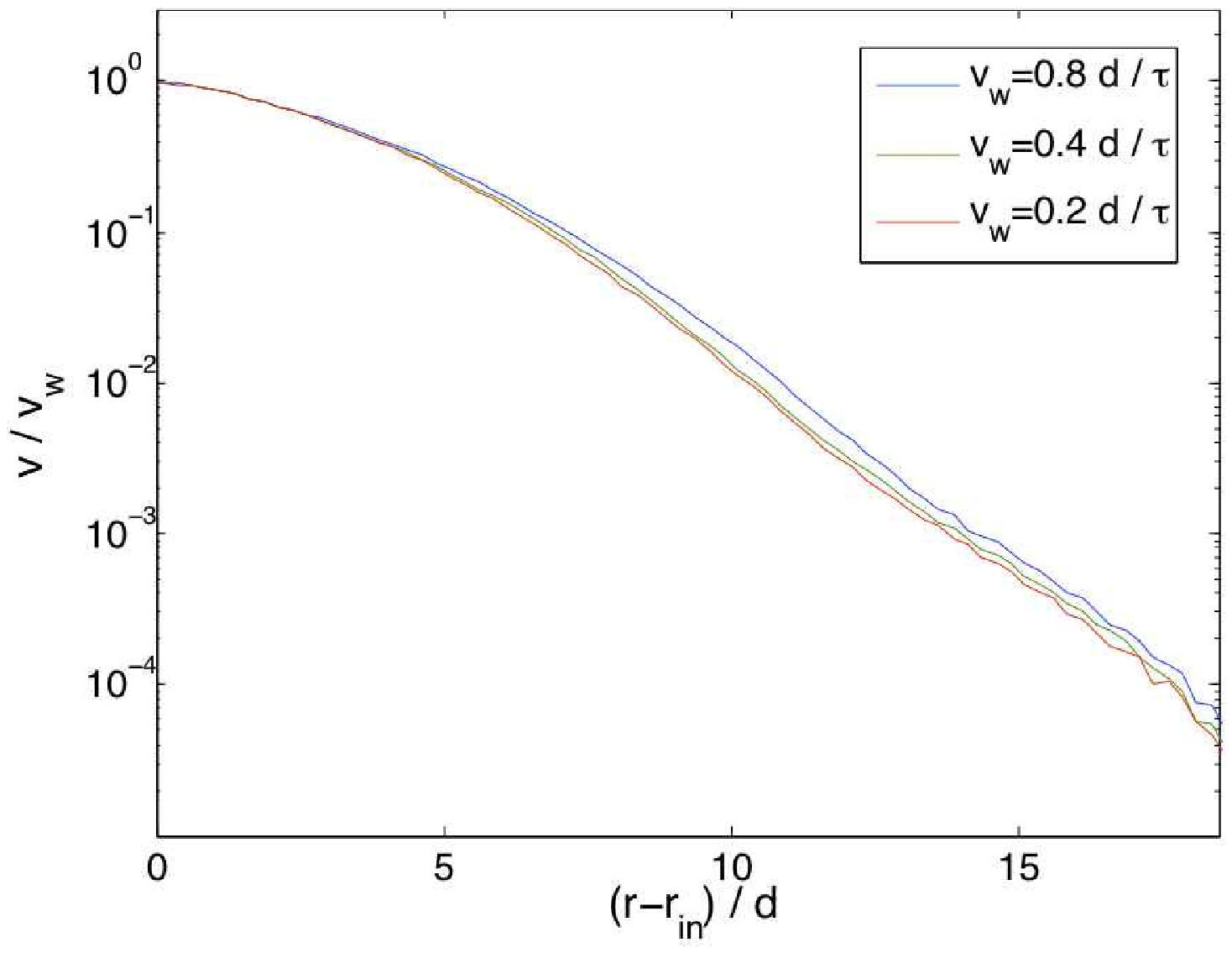, width=3in}
  \caption{Velocity profiles for three different angular velocities, with
  $\mu_c=0.5$ and $\rin=40d$. The time windows over which the velocities are
  computed are scaled according to the angular velocity.}
  \label{fig:profiles_vel}
\end{figure}

In agreement with past work on Couette flow \cite{losert00,
  bocquet01}, we too find that the normalized flow profile is roughly
unaffected by the shearing rate (see Figure
\ref{fig:profiles_vel}). As previously discussed, this behavior is in
agreement with the SFR, which always permits flow fields to be
multiplied by a constant.


\begin{table}
  \centering
  \begin{tabular}{c|c}
    $\mu_c$ & $b$ \\
    \hline
    $0.1$ & $0.974d$ \\
    $0.3$ & $0.983d$ \\
    $0.5$ & $1.033d$ \\
    $0.7$ & $1.046d$ \\
    $0.9$ & $1.032d$ \\
  \end{tabular} \qquad
  \begin{tabular}{c|c}
    $\phi$ & $b$\\
    \hline
    $20^\circ$ & $1.026d$\\
    $22^\circ$ & $1.038d$\\
    $24^\circ$ & $1.052d$\\
    $26^\circ$ & $1.069d$\\
    $28^\circ$ & $1.084d$\\
    $30^\circ$ & $1.102d$\\
  \end{tabular}
  \caption{(Left) Simulation: Half-widths of the shearing velocity profiles for
  different values of $\mu_c$, calculated by fitting the functional form
  $f(r)=a-(\log 2)r/b$ to $\log v/\vw$ over the range $45d<r<58d$. (Right)
  SFR: Fits the predictions to the same form and uses $L_s=3d$.}
  \label{tab:halfwidths}
\end{table}

\section{Conclusion}

We have presented the crucial principles motivating the Stochastic
Flow Rule and have assessed its validity by checking analytical
predictions for silo and annular Couette flow against discrete-element
simulations.  Using the same parameters for both cases, without any
fitting, the SFR manages good predictions for these two very different
flow geometries, which it seems cannot be described, even
qualitatively, by any other model. The model was also shown to capture
the ``diffusive'' type flow properties unique to granular materials
such as Gaussian downward velocity in the draining silo and
exponentially decaying velocity in the annular Couette cell.

Our simulations also indicate that the slight changes in flow brought
on by varying the inter-particle contact friction in the annular flow
geometry match the trends the SFR predicts when the internal friction
angle is approrpiately varied. The trend is also captured when the
inner wall radius is varied, though the quantitative agreement is not
as strong. In agreement with past studies on annular Couette flow, and
in validation of one of the first principles behind the SFR, we find
in our simulations that the flow rate does not significantly affect
the normalized flow profile over a significant range of rates.

These results, taken together with our prior work on spot-driven
particle simulations~\cite{rycroft06a}, suggest that the SFR provides
a general paradigm for multiscale modeling of granular flow or, more
generally perhaps, plastic deformation of other amorphous
materials. Whenever the stress state is near yielding our physical
picture of flow is that spots of fluidization perform random walks
along slip-lines biased by local stress imbalances.  With some
exceptions as outlined in \cite{kamrin06} (so-called ``slip-line
admissible'' geometries) we believe the SFR is a highly general law
for 2D dense granular flow.

\bibliographystyle{plain}
\bibliography{granular}

\end{document}